\PassOptionsToPackage{utf8}{inputenc}
\documentclass[nocrop,letter]{bioinfoed}

\usepackage{natbib}
\usepackage{graphicx}
\usepackage{algorithm,algorithmic}
\newcommand{\ts}{\textsuperscript}
\usepackage{amsmath}
\usepackage{amsthm}
\usepackage{amssymb}
\usepackage{siunitx}
\usepackage{float}
\usepackage{booktabs}
\usepackage{bm}
\usepackage{todonotes}

\usepackage{caption}
\usepackage{subcaption}
\usepackage{soul}
\usepackage{cancel}
\usepackage{hyperref}

\newcommand{\red}[1]{\textcolor{black}{#1}} 

\begin{document}
\firstpage{1}


\title[Drug Combo Generator]{Network-principled deep generative models for designing drug combinations as graph sets}
\author[Karimi, Hasanzadeh and Shen]{Mostafa Karimi\,$^{\text{\sfb 1,2,=}}$, Arman Hasanzadeh\,$^{\text{\sfb 1,=}}$ and Yang shen\,$^{\text{\sfb 1,2,*}}$}
\address{$^{\text{\sf 1}}$Department of Electrical and Computer Engineering and $^{\text{\sf 2}}$TEES--AgriLife Center for Bioinformatics and Genomic Systems Engineering, Texas A\&M University, College Station, 77843, USA.}

\corresp{$^=$Co-first authors.\\$^\ast$To whom correspondence should be addressed.}

\history{Received on XXXXX; revised on XXXXX; accepted on XXXXX}

\editor{Associate Editor: XXXXXXX}

\abstract{\textbf{Motivation:}
Combination therapy has shown to improve therapeutic efficacy while reducing side effects.  Importantly, it has become an indispensable strategy to overcome resistance in antibiotics, anti-microbials, and anti-cancer drugs.  Facing enormous chemical space and unclear design principles for small-molecule combinations, computational drug-combination design has not seen generative models to meet its potential to accelerate resistance-overcoming drug combination discovery. \\
\textbf{Results:} We have developed the first deep generative model for drug combination design, by jointly embedding graph-structured domain knowledge and iteratively \red{training} a reinforcement learning-based chemical graph-set designer.  First, we have developed Hierarchical Variational Graph Auto-Encoders (HVGAE) trained end-to-end to jointly embed gene-gene, gene-disease, and disease-disease networks.  Novel attentional pooling is introduced here for learning disease-representations from associated genes' representations.  Second, targeting diseases in learned representations, we have recast the drug-combination design problem as graph-set generation and developed a deep learning-based model with novel rewards.  Specifically, besides chemical validity rewards, we have introduced novel generative adversarial award, being generalized sliced Wasserstein, for chemically diverse molecules \red{with distributions similar} to known drugs.  We have also designed a network principle-based reward for drug combinations.  Numerical results indicate that, compared to state-of-the-art graph embedding methods, HVGAE learns more informative and generalizable disease representations. Results also show that the deep generative models generate drug combinations following the principle across diseases.  Case studies on \red{four diseases show that network-principled drug combinations tend to have low toxicity.}   The generated drug combinations collectively cover the disease module similar to FDA-approved drug combinations and could potentially suggest novel systems-pharmacology strategies.  Our method allows for examining and following network-based 
principle or hypothesis to efficiently generate disease-specific drug
combinations in a vast chemical combinatorial space.  \\
\textbf{Availability: \url{https://github.com/Shen-Lab/Drug-Combo-Generator}} \\
\textbf{Contact:} \href{yshen@tamu.edu}{yshen@tamu.edu}\\
\textbf{Supplementary information}: Supplementary data are available at \\ \url{https://github.com/Shen-Lab/Drug-Combo-Generator/blob/master/SI_drugcomb_RL.pdf}
}

\maketitle

\section{Introduction}
Drug resistance is a fundamental barrier to developing robust  antimicrobial and anticancer therapies~\citep{taubes2008bacteria,housman2014drug}.  Its first sign was observed in 1940s soon after the discovery of penicillin~\citep{abraham1940enzyme}, the first modern antibiotic.  Since then, drug resistance has surfaced and progressed in infectious diseases such as HIV ~\citep{clavel2004hiv}, tuberculosis (TB)~\citep{dooley1992multidrug} and hepatitis~\citep{ghany2007drug} as well as cancers~\citep{holohan2013cancer}. Mechanistically, it can emerge through drug efflux \citep{chang2001structure}, activation of alternative pathways \citep{lovly2014molecular} and protein mutations \citep{toy2013esr1,balbas2013overcoming} while decreasing the efficacy of drugs.

Combination therapy is a resistance-overcoming strategy that has found success in combating HIV \citep{shafer1999highly}, TB \citep{ramon2011synergistic}, cancers  \citep{sharma2015immune,bozic2013evolutionary} and so on. Considering that most diseases and their resistances are multifactorial \citep{kaplan2000genomics,keith2005multicomponent}, multiple drugs targeting multiple components simultaneously could confer less resistance than individual drugs targeting components separately.  Examples include targeting both MEK and BRAF in patients with BRAF V600-mutant melanoma rather than targeting 
MEK or BRAF alone \citep{madani2018predictive,flaherty2012combined}. The effect of drug combination is usually categorized as synergistic, additive, or antagonistic depending on whether it is greater than, equal to or less than the sum of individual drug effects \citep{chou2006theoretical}. Synergistic combinations are effective at delaying the beginning of the resistance, however antagonistic combinations are effective at suppressing expansion of resistance \citep{saputra2018combination,singh2017suppressive}, representing offensive and defensive strategies to overcome drug resistance.  
In particular, offensive strategies cause huge early causalities but defensive ones anticipate and develop protection against future threats. \citep{saputra2018combination}.

Discovering a drug combination to overcome resistance is however extremely challenging, even more so than discovering a drug which is already a costly ($\sim$billions of USD) \citep{dimasi2016innovation} and lengthy ($\sim$12 years) \citep{van2016drugs} process with low success rates (3.4\% phase-1 oncology compounds make it to approval and market) \citep{wong2019estimation}. An apparent challenge, a combinatorial one, is in the scale of chemical space, which is estimated to be $10^{60}$ for single compounds \citep{chemspace96} and can ``explode'' to $10^{60K}$ for $K$-compound combinations. Even if the space is restricted to around $10^3$ FDA-approved human drugs, there are  $10^5$--$10^6$ pairwise combinations.  Another challenge, a conceptual one, is in the complexity of systems biology.  On top of on-target efficacy and off-target side effects or even toxicity that need to be considered for individual drugs, network-based design principles are much needed for drug combinations that effectively target multiple proteins in a disease module and have low toxicity or even resistance profiles \citep{martinez2016should,billur2014network}.

Current computational models in drug discovery, especially those for predicting pharmacokinetic and pharmacodynamic properties of individual drugs/compounds, can be categorized into discriminative and generative models. Discriminative models predict the distribution of a property for a given molecule whereas generative models would learn the joint distribution on the property and molecules.  
For instance, discriminative models have been developed for predicting single compounds' toxicities, based on support vector machines \citep{darnag2010support}, random forest \citep{svetnik2003random} and deep learning \citep{mayr2016deeptox}. Whereas discriminative models are useful for evaluating given compounds or even searching compound libraries, generative models can effectively design compounds of desired properties in chemical space. Recent advance in inverse molecular design has seen deep generative models such as SMILES representation-based reinforcement learning  \citep{popova2018deep} or recurrent neural networks (RNNs) as well as graph representation-based generative adversarial network (GANs), reinforcement learning  \citep{you2018graph}, and generative tensorial reinforcement learning (GENTRL) \citep{zhavoronkov2019deep}. 

Unlike single drug design, current computational efforts for drug combinations are exclusively focused on  discriminative models and lack generative models. The main focus for drug combination is to use discriminate models to identify synergistic or antagonistic drugs for a given specific disease. Examples include the  Chou-Talalay method \citep{chou2010drug}, integer linear programming \citep{pang2014combinatorial}, and deep learning \citep{preuer2017deepsynergy} and . However, it is daunting if not infeasible to enumerate all cases in the enormous chemical combinatorial space and evaluate their combination effects using a discriminative model.  Not to mention that such methods often lack explainability.

Directly addressing aforementioned combinatorial and conceptual challenges and filling the void of generative models for drug combinations, in this study, we develop network-based representation learning for diseases and deep generative models for accelerated and principled drug combination design \red{(the general case of $K$ drugs)}.  Recently, by analyzing the network-based relationships between disease proteins and drug targets in the human protein–protein interactome, \citeauthor{cheng2019network} proposed an elegant principle for FDA-approved drug combinations that targets of two drugs both hit the disease module but cover different neighborhoods.  Our methods allow for examining and following the proposed network-based principle \citep{cheng2019network} to efficiently generate disease-specific drug combinations in a vast chemical combinatorial space.  They will also help meet a critical need of computational tools in a  battle against quickly evolving bacterial, viral and tumor populations with accumulating resistance.

To tackle the problem, we have developed a network principle-based deep generative model for faster, broader and deeper exploration of drug combination space by following the principle underling FDA approved drug combinations. First, we have developed Hierarchical Variational Graph Auto-Encoders (HVGAE) for jointly embedding disease-disease network and gene-gene network. 
Through end-to-end training, we embed genes in a way that they can represent the human interactome. Then, we utilize their embeddings with novel attentional pooling to create features for each disease so that we can embed diseases more accurately. Second, we have also developed a reinforcement-learning based graph-set generator for drug combination design by utilizing both gene/disease embedding and network principles. Besides those for chemical validity and properties, our rewards also include 1) a novel adversarial reward, generalized sliced Wasserstein distance, that fosters generated molecules to be diverse yet similar in distribution to known compounds (ZINC database and FDA-approved drugs) and 2) a network principle-based reward for drug combinations that are feasible for online calculations.  

\begin{figure*}[!htb]
    \centering
    \includegraphics[width=0.9\textwidth, keepaspectratio]{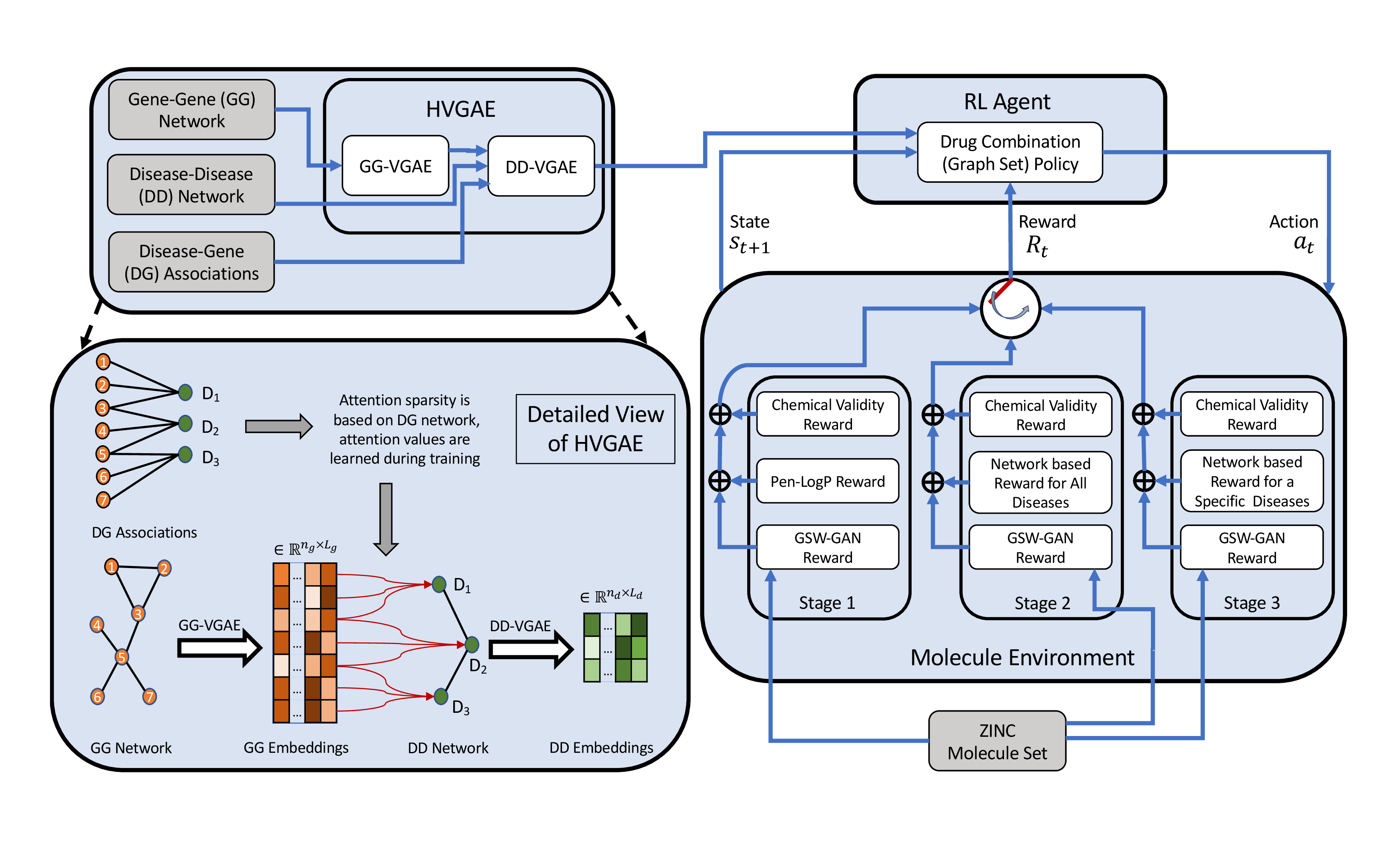}
    \caption{\red{Overall schematics of the proposed approach for generating  disease-specific drug combinations.}}
    \label{fig:scheme}
\end{figure*}

\red{The overall schematics are shown in Fig.~\ref{fig:scheme} and details in Sec.~\ref{sec:methods}.}

\section{Data}

\subsection{Human interactome and its features}
We used the human interactome data (a gene-gene network) from \citep{menche2015uncovering} that feature 13,460 proteins interconnected by 141,296 interactions.  

We introduced edge features for the human interactome based on the biological nature of edges (interactions).  The interactome was compiled by combining experimental support from various sources/databases including 1) regulatory interactions from TRANSFAC \citep{matys2003transfac}; 2) binary interactions from high-throughput (including \citep{rolland2014proteome}) and literature-curated datasets (including IntAct \citep{aranda2010intact} and MINT \citep{ceol2010mint}) as well as literature-curated interactions from low-throughput experiments (IntAct, MINT, BioGRID \citep{stark2010biogrid}, and HPRD \citep{keshava2009human}); 3) metabolic enzyme-coupled interactions from \citep{lee2008implications}; 4) protein complexes from CORUM \citep{ruepp2010corum}; 5) kinase-substrate pairs from PhosphositePlus \citep{hornbeck2012phosphositeplus}; and 6) signaling interactions.  In summary, an edge could correspond to one or  multiple physical interaction types. So we used a 6-hot encoding for edge features, based on whether an edge corresponds to regulatory, binary, metabolic, complex, kinase and signaling interactions.

We also introduced features for nodes (genes) in the human interactome based on 1) KEGG pathways \citep{kanehisa2002kegg} (336 features) queried through Biopython \citep{cock2009biopython}; 2) Gene Ontology (GO) terms \citep{ashburner2000gene} including biological process (30,769 features), molecular function (12,183 features), and cellular component (4,451 features), mapped using the NCBI Gene2Go dataset; 3) disease-gene associations from the database OMIM (Mendelian Inheritance in Man)  \citep{hamosh2005online} and the results from Genome-Wide Association Studies (GWAS)  \citep{mottaz2008mapping,ramos2014phenotype} (299 features). The last 299 features correspond to 299 diseases represented by the Medical Subject Headings (MeSH) vocabulary \citep{rogers1963medical}. 

After removing those genes without KEGG pathway information, the human interactome used in this study has 13,119 genes and 352,464 physical interactions.

\subsection{Disease-disease network}

We used a disease-disease network from \citep{menche2015uncovering} with 299 nodes (diseases),  created based on human interactome data (as detailed earlier), gene expression data \citep{su2004gene}, disease-gene associations \citep{mottaz2008mapping,ramos2014phenotype,hamosh2005online}, Gene Ontology \citep{ashburner2000gene}, symptom similarity \citep{zhou2014human} and comorbidity \citep{hidalgo2009dynamic}. The original disease-disease network is a complete graph with real-valued edges. The edge value between two diseases shows how much they are topologically separated from each other. A positive/negative edge weight indicates that that two disease modules are topologically separated/overlapped. Therefore, we used zero-weight as the threshold and pruned positive-valued edges, which results in a disease-disease network of 299 nodes and 5,986 edges (without weights).

\subsection{Disease-gene associations}
We used disease-gene associations from the database OMIM  \citep{hamosh2005online}.  These associations bridge aforementioned gene-gene and disease-disease networks into a hierarchical graph of genes and diseases, based on which gene and disease representations will be learned.   

\subsection{Disease classification}\label{sec:diseaseClass}
For the purpose of assessment, we used the Comparative Toxicogenomics Database (CTD) \citep{davis2019comparative} to classify diseases into 8 classes based on their Disease Ontology (DO) terms \citep{schriml2012disease} where diseases are  represented in the MeSH vocabulary \citep{rogers1963medical}.  In the CTD database only 201 of the 299 diseases have a corresponding DO term. Therefore, for the 98 diseases with missing DO terms we considered the majority of their parents' DO terms, if applicable, as their DO terms. With this approach, we assigned DO terms to 66 such diseases and classified 267 of the 299 diseases.  The 
32 disease with DO terms still missing are usually at the top layers of the MeSH tree.

\subsection{FDA-approved drugs and drug combinations}

To assess our deep generative model for drug combination design (to be detailed in Sec. 3.2), we consider a comprehensive list of US FDA-approved combination drugs (1940--2018.9) \citep{das2018survey}.  The dataset contains 419 drug combinations consisting of 328 unique drugs, including 341 (81\%), 67 (16\%) and 11 (3\%) of double, triple and quadruple drug combinations. 

We also utilized the  curated drug-disease association from CTD database \citep{davis2019comparative}.

\section{Methods}\label{sec:methods}

We have developed a network-based drug combination generator which can be utilized in overcoming drug resistance. Representing drugs through their molecular graphs, we recast the problem of drug combination generation into  network-principled, graph-set generation by incorporating prior knowledge such as human interactome (gene-gene), disease-gene, disease-disease, gene pathway, and gene-GO  relationships. Furthermore, we formulate the graph-set generation problem as learning a Reinforcement Learning (RL) agent that iteratively adds substructures and edges to each molecular graph in a chemistry- and system-aware environment. 
To that end, the RL model is trained to maximize a desired property $Q$ (for example, therapeutic efficacy for drug combinations) while following the valency (chemical validity) rules and being similar in distribution to the prior set of graphs.

\red{As shown in Fig.~\ref{fig:scheme},} the proposed approach consists of: 1) embedding prior knowledge (different network relationships) through Hierarchical Variational Graph Auto-Encoders (HVGAE); and 2) generating drug combinations as graph sets through a reinforcement learning algorithm, which will be detailed next. 

\paragraph{Notations:} As both gene-gene and disease-disease networks can be represented as graphs, notations are differentiated by superscripts `g' and `d' to indicate gene-gene and disease-disease networks, respectively.  Drugs (compounds) are also represented as graphs and notations with `$k$' in the superscript indicates the $k$-th drug (graph) in the drug combination (graph set).  

\subsection{Hierarchical Variational Graph Auto-Encoders (HVGAE) for representation learning}\label{sec: hvgae}

Suppose that a gene-gene network is represented as a graph $G^{(\mathrm{g})}=(A^{(\mathrm{g})},\{F^{(\mathrm{g},m)}\}_{m=1}^{M})$, where
$A^{(\mathrm{g})} = [A^{(\mathrm{g}, 1)}, \cdots, A^{(\mathrm{g}, n_e)}] \in \{0,1\}^{n_\mathrm{g}\times n_\mathrm{g} \times n_e}$ 
is the adjacency tensor of the gene-gene network with $n_\mathrm{g}$ nodes and $n_e$ edge types ($k$-hot encoding of 6 types of aforementioned physical interactions such as regulatory, binary, metabolic, complex, kinase and signaling interactions). 
We also define $\Tilde{A}^{(\mathrm{g})} \in \{0,1\}^{n_\mathrm{g}\times n_\mathrm{g}}$ to be elemenwise $\mathrm{OR}$ of $\{A^{(\mathrm{g}, 1)}, \cdots, A^{(\mathrm{g}, n_e)}\}$. 
Furthermore, $F^{(\mathrm{g},m)}$ denotes the $m$\ts{th} set of node features for gene-gene network where $M$ (5 in the study) represents different types of node features such as pathways, 3 GO terms and gene-disease relationship. We also suppose the disease-disease network is represented as graph $G^{(\mathrm{d})}=(A^{(\mathrm{d})},F^{(\mathrm{d})})$, where $A^{(\mathrm{d})} \in \{0,1\}^{n_\mathrm{d}\times n_\mathrm{d}}$ is the adjacency matrix of the disease-disease network with $n_\mathrm{d}$ nodes; and $F^{(\mathrm{d})}$ represents the set of node features for the  disease-disease network. 

\red{W}e have developed a hierarchical embedding with 2 levels. In the first level, we embed the gene-gene network to get the features related to each disease and then we incorporate the disease features within the disease-disease network to embed their relationship. We infer the embedding for each gene and disease jointly through end-to-end training. The proposed HVGAE perform probabilistic auto-encoding to capture uncertainty of representations which is in the same spirit as the variational graph auto-encoder models introduced in \citep{kipf2016variational,hasanzadeh2019semi,hajiramezanali2019variational} . 

\subsubsection{First level: Gene-Gene embedding}
The inference model for variational embedding of the gene-gene network is formulated as follows.  We first use $M$ 
graph neural networks (GNNs) to transform individual nodes' features in $M$ types and then concatenate the $M$ sets of results $\hat{F}^{(\mathrm{g},m)}$ ($m=1,\ldots,M$) into $\hat{F}^{(\mathrm{g})}$:  

\begin{gather}
    \hat{F}^{(\mathrm{g},m)} = \mathrm{AGG}\left(\{\mathrm{GNN}_j(A^{(\mathrm{g},j)},F^{(\mathrm{g},m)})\},\, j=1,\cdots, n_e \right) \nonumber\\
    \hat{F}^{(\mathrm{g},m)} \in \mathbb{R}^{n_\mathrm{g} \times L_\mathrm{g}}, \quad m=1,\cdots,M \nonumber\\
    \hat{F}^{(\mathrm{g})} = \text{CONCAT}(\{\hat{F}^{(\mathrm{g},m)}\}_{m=1}^{M}) \in \mathbb{R}^{n_\mathrm{g} \times M L_\mathrm{g}},
\end{gather}
where $\mathrm{AGG}$ is an aggregation function combining output features of $\mathrm{GNN}_j$'s for each node. 
We used a two layer fully connected neural network with ReLU activation functions followed by a single linear layer in our implementation. 
We then approximate the posterior distribution of stochastic latent variables $\bm{Z}^{(\mathrm{g})}$ (containing $\bm{z}_i^{(\mathrm{g)}} \in \mathbb{R}^{L_\mathrm{g}}$ for $i=1,\cdots, n_\mathrm{g}$ where $L_\mathrm{g}$ (32 in this study)
is the latent space dimensionality for the $i$\ts{th} gene), with a multivariate Gaussian distribution $q(\cdot)$ given the gene-gene network's aggregated node features $\hat{F}^{(\mathrm{g})}$ and adjacency tensor $A^{(\mathrm{g})}$: 

\begin{gather}
    q(\bm{Z}^{(\mathrm{g})}|\hat{F}^{(\mathrm{g})},A^{(\mathrm{g})}) = \prod_{i=1}^{n_\mathrm{g}} q(\bm{z}_i^{(\mathrm{g})} | \hat{F}^{(\mathrm{g})},A^{(\mathrm{g})}), \text{where} \nonumber\\
    q(\bm{z}_i^{(\mathrm{g})} | \hat{F}^{(\mathrm{g})},A^{(\mathrm{g})}) = \mathcal{N}(\bm{\mu}_i^{(\mathrm{g})},\textit{diag}(\bm{\sigma}_i^{2,(g)})), \nonumber\\
    \bm{\mu}^{(\mathrm{g})} = \mathrm{AGG}\left(\{\mathrm{GNN}_{\bm{\mu},g,j}(A^{(\mathrm{g},j)},\hat{F}^{(\mathrm{g})})\},\, j=1,\cdots, n_e \right), \nonumber\\
    \log(\bm{\sigma}^{(\mathrm{g})}) = \mathrm{AGG}\left(\{\mathrm{GNN}_{\bm{\sigma},g,j}(A^{(\mathrm{g},j)},\hat{F}^{(\mathrm{g})})\},\, j=1,\cdots, n_e \right), \nonumber\\
    \bm{\mu}^{(\mathrm{g})} \in \mathbb{R}^{n_\mathrm{g} \times L_\mathrm{g}}, \quad \log(\bm{\sigma}^{(\mathrm{g})}) \in \mathbb{R}^{n_\mathrm{g} \times L_\mathrm{g}}.
\end{gather}

where $\bm{Z}^{(\mathrm{g})} \in \mathbb{R}^{n_\mathrm{g} \times L_\mathrm{g}}$; $\bm{\mu}^{(\mathrm{g})}$ is the matrix of mean vectors $\bm{\mu}_i^{(\mathrm{g})}$; and $\bm{\sigma}^{(\mathrm{g})}$ the matrix of standard deviation vectors $\bm{\sigma}_i^{(\mathrm{g})}$ ($i=1,\ldots,n_\mathrm{g}$). 

The generative model for the gene-gene network is formulated as:
    \begin{gather}
        p(\Tilde{A}^{(\mathrm{g})}|\bm{Z}^{(\mathrm{g})}) = \prod_{i=1}^{n}\prod_{j=1}^{n}\,  p(\Tilde{A}^{(\mathrm{g})}_{ij}|\bm{z}^{(\mathrm{g})}_i,\bm{z}^{(\mathrm{g})}_j), \text{where} \nonumber\\
        p(\Tilde{A}^{(\mathrm{g})}_{ij}|\bm{z}^{(\mathrm{g})}_i,\bm{z}^{(\mathrm{g})}_j) = \sigma(\bm{z}^{(\mathrm{g})}_i \bm{z}^{(\mathrm{g}) T}_j),
    \end{gather}
and $\sigma(\cdot)$ is the logistic sigmoid function. 
The loss for gene-gene variational embedding is represented as a variational lower bound (ELBO):
\begin{equation}
    \begin{split}
        \mathcal{L}^{(\mathrm{g})} =& \, \mathbb{E}_{q(\bm{Z}^{(\mathrm{g})}|\hat{F}^{(\mathrm{g})},A^{(\mathrm{g})})}[\log p(\Tilde{A}^{(\mathrm{g})}|\bm{Z}^{(\mathrm{g})})] \\
        &- \mathrm{KL}\big(q(\bm{Z}^{(\mathrm{g})}|\hat{F}^{(\mathrm{g})},A^{(\mathrm{g})})||p(\bm{Z}^{(\mathrm{g})})\big),
    \end{split}
\end{equation}
where $\mathrm{KL}\big(q(\cdot)||p(\cdot)\big)$ is the Kullback-Leibler divergence between $q(\cdot)$ and $p(\cdot)$. We take the Gaussian prior for $p(\bm{Z}^{(\mathrm{g})})$ and make use of the reparameterization trick \citep{kipf2016variational} for training. 

\subsubsection{Second level: disease-disease embedding}

The inference model for variational embedding of the disease-disease network is similar to that of the gene-gene network except that the disease-disease network's aggregated node features, $\hat{F}^{(\mathrm{d})}$, are derived through parameterized attentional pooling of $\hat{\bm{Z}}^{(\mathrm{g})}_r$,  latent variables of genes associated with  the $r$\ts{th} disease (a subset of $\bm{Z}^{(\mathrm{g})}$): 
    \begin{gather}
        \bm{e}_r = \bm{v}\,\, \mathrm{tanh}(\hat{\bm{Z}}^{(\mathrm{g})}_r W + \bm{b}), \quad  r=1,\cdots,n_\mathrm{d} \nonumber\\
        \bm{\alpha}_r = \text{softmax}(\bm{e}_r), \quad  r=1,\cdots,n_\mathrm{d} \nonumber\\
        \hat{F}_r^{(\mathrm{d})} = \sum_{i}  \bm{\alpha}_{r,i} \hat{\bm{Z}}^{(\mathrm{g})}_{r,i}, \quad r=1,\cdots,n_\mathrm{d} \nonumber\\
        \hat{F}^{(\mathrm{d})} = \text{CONCAT}(\{\hat{F}_r^{(\mathrm{d})}\}_{r=1}^{n_\mathrm{d}}) \in \mathbb{R}^{n_\mathrm{d} \times L_\mathrm{d}},
        \end{gather}
where  $\bm{\alpha}_m$ capture the importance of genes related to the $r$\ts{th} disease for calculating its latent representations and $L_\mathrm{d}$ is the latent space dimensionality of a disease. 

Once $\hat{F}^{(\mathrm{d})}$, the disease-disease network's aggregated node features for all diseases, are derived; we again define   $q(\bm{Z}^{(\mathrm{d})}|\hat{F}^{(\mathrm{d})},A^{(\mathrm{d})})$ for the posterior distribution of stochastic latent variables $\bm{Z}^{(\mathrm{d})}$ similarly to what we did in Eq. (2) except that AGG functions are removed since disease-disease network has one binary adjacency matrix; give the generative decoder $p(A^{(\mathrm{d})}|\bm{Z}^{(\mathrm{d})})$ for embedding the disease-disease network similarly to what we did in Eq. (3); and calculate the variational lowerbound (ELBO) loss $\mathcal{L}^{(\mathrm{d})}$ for the disease-disease network similarly to what we did in Eq. (4). Details can be found in Supplemental Sec. 1.1.

Both levels of our proposed HVGAE, i.e. gene-gene and disease-disease variational graph representation learning, are jointly trained in an end-to-end fashion using the following overall loss:
\begin{equation}
    \mathcal{L}^{\text{HVGAE}} = \mathcal{L}^{(\mathrm{d})} + \mathcal{L}^{(\mathrm{g})}.
\end{equation}

\subsection{Reinforcement learning-based graph-set generator for drug combinations}
In this section, we introduce the reinforcement learning-based drug combination generator. We will detail 1) the state space of graph sets \red{($K$ compounds)} and the action space of graph\red{-set} growth; 2) multi-objective rewards including chemical validity and our generalized sliced Wasserstein reward for individual drugs as well as our newly designed network principle-based reward for drug combinations; and 3) policy network that learns to take actions in the rewarding environment.  

\subsubsection{State and action space}
We represent a graph set (drug combination) with $K$ graphs as $\mathcal{G} = \{ G^{(k)}\}_{k=1}^{K}$. Each graph  $G^{(k)}=(A^{(k)},E^{(k)},F^{(k)})$ where $A^{(k)} \in \{0,1\}^{n_{k}\times n_{k}}$ is the adjacency matrix, $F^{(k)} \in \mathbb{R}^{n_{k}\times \phi}$ the node feature matrix, $E^{(k)} \in \{0,1\}^{\epsilon\times n_{k}\times n_{k}}$ the edge-conditioned adjacency tensor, and $n_{k}$ the number of vertices for the $k$\ts{th} graph, respectively; and $\phi$ is the number of features per nodes and $\epsilon$ the number of edge types.

The state space  $\mathcal{G}$ is the set of all $K$ graphs with different numbers and types of nodes or  edges. Specifically, the state of the environment $s_t$ at iteration $t$ is defined as the intermediate graph set  $\mathcal{G}_t = \{ G^{(k)}_t\}_{k=1}^{K}$ generated so far which is fully observable by the RL agent.

The action space is the set of edges that can be added to the graph set. An action $a_t$ at iteration $t$ is analogous to link prediction in each graph in the set. More specifically, a link can either connect a new subgraph (a single node/atom or a subgraph/drug-substructure) to a node in $G_t^{(k)}$ or connect existing nodes within graph $G_t^{(k)}$. The actions can be interpreted as connecting the current graph with a member of scaffold subgraphs set $C$. Mathematically,  for $G_t^{(k)}$, graph $k$ at step $t$, the action $a_t^{(k)}$ is the quadruple of $a_t^{(k)} = \textsc{concat}(a_{\mathrm{first, t}}^{(k)},a_{\mathrm{second, t}}^{(k)},a_{\mathrm{edge, t}}^{(k)},a_{\mathrm{stop, t}}^{(k)})$.

\subsubsection{Multi-objective reward}
We have defined a multi-objective reward $R_t$ to satisfy certain requirements in drug combination therapy.   First, a chemical validity reward maintains that individual compounds are chemically valid.  Second, a novel adversarial reward, generalized sliced Wasserstein GAN (GS-WGAN), enforces generated compounds are synthesizable and ``drug-like" by following the distribution of synthesizable compounds in the ZINC database \citep{irwin2005zinc} or FDA-approved drugs.  Third, a network principle-based award would encourage individual drugs to target the desired disease module but not to overlap in their target sets. \red{Toxicity due to drug-drug interactions can also be included as a reward. It is intentionally left out in this study so that toxicity can be evaluated for drug combinations designed to follow the network principle.}

When training the RL agent, we use different reward combinations in different stages.  We first only use the weighted combination of chemical validity and GS-WGAN awards learning over drug combinations for all diseases; then we remove the \red{penalized logP (Pen-logP)}  portion of chemical validity and add adversarial loss again while learning over drug combinations for all diseases; and finally use the combination of the three rewards as in the second stage but focusing on a target disease and possibly on restricted actions/scaffolds (in a spirit similar to transfer learning).  The three types of rewards are detailed as follows.

\noindent\textbf{Chemical validity reward for individual drugs.}\quad A small positive  reward is assigned if the action does not violate valency rules. Otherwise a small negative reward is assigned. This is an intermediate reward added at each step. Another reward is on \red{penalized} logP (lipophilicity where P is the octanol-water partition coefficient) or Pen-logP values.  The design and the parameters of this reward is adopted from \citep{you2018graph} without optimization.  

\noindent\textbf{Adversarial reward using generalized sliced Wasserstein distance (GSWD).}\quad To ensure that the generated molecules resemble a given set of molecules (such as those in ZINC or FDA-approved), we deploy Generative Adversarial Networks (GAN). GANs are very successful at modeling high-dimensional distributions from given samples. However they are known to suffer from training unsuitability and cannot generate diverse samples (a phenomenon known as \textit{mode collapse}). 

Wasserstein GANs (WGAN) have shown to improve stability and mode collapse by replacing the Jenson-Shannon divergence in original GAN formulation with the Wasserstein Distance (WD) \citep{arjovsky2017wasserstein}. More specifically, the objective function in WGAN with gradient penalty \citep{gulrajani2017improved} is defined as follows:
\begin{gather}\label{equ: wgan}
    \min_{\theta} \max_{\phi} V_\mathrm{W}(\pi_{\theta}, D_{\phi}) + \lambda R(D_{\phi}),\\
    \text{with}\quad V_W(\pi_{\theta}, D_{\phi}) = \mathbb{E}_{\mathbf{x}\sim p_{r}}[\log D_{\phi}(\mathbf{x})] -
    \mathbb{E}_{\mathbf{y}\sim \pi_{\theta}}[\log D_{\phi}(\mathbf{y})],\nonumber
\end{gather}
where $p_r$ is the data distribution, $\lambda$ is a hyper-parameter, $R$ is the Lipschitz continuity regularization term, $D_{\phi}$ is the critic with parameters $\phi$, and $\pi_{\theta}$ is the policy (generator) with parameters $\theta$.

Despite theoretical advantages of WGANs, solving equation \eqref{equ: wgan} is computationally expensive and intractable for high dimensional data. To overcome this problem, we propose and formulate a novel Generalized Sliced WGAN (GS-WGAN) which deploys Generalized Sliced Wasserstein Distance (GSWD) \citep{kolouri2019generalized}. GSWD, first, factorizes high-dimensional probabilities into multiple marginal 1D distributions with generalized Radon transform. Then, by taking advantage of closed form solution of Wasserstein distance in 1D, the distance between two distributions is approximated by the sum of Wasserstein distances of marginal 1D distributions.
More specifically, let $\mathcal{R}$ represent generalized Radon transform operator. The generalized Radon transform (GRT) of a probability distribution $\mathbb{P}(\cdot)$ which is defined as follows:
\begin{gather}\label{equ: radon}
    \mathcal{R}P(t, \psi) = \int_{\mathbb{R}^d} \mathbb{P}(\mathbf{x})\, \delta(t - f(\mathbf{x}, \psi)) \, d\mathbf{x},
\end{gather}
where $\delta(\cdot)$ is the one-dimensional Dirac delta function, $t \in \mathbb{R}$ is a scalar, $\psi$ is a unit vector in the unit hyper-sphere in a $d$-dimensional space ($\mathbb{S}^{d-1}$), and $f$ is a  projection function whose parameters will be learned in training. Injectivity of the GRT \citep{beylkin1984inversion}  is the requirement for the GSWD to be a valid distance. 
We use linear project $f(x,\psi)$ here and can easily extend to two nonlinear cases that maintains the GRT-injectivity (circular nonlinear projections or homogeneous polynomials with an odd degree).  

GSWD between two d-dimensional distributions $\mathbb{P}_{X}$ and $\mathbb{P}_{Y}$ is therefore defined as:
\begin{gather}\label{equ: gswd}
    \text{GSWD}(\mathbb{P}_{X},\, \mathbb{P}_{Y}) = \int_{\mathbb{S}^{d-1}} \text{WD} (\mathcal{R}P_{X}(\cdot , \psi),\, \mathcal{R}P_{Y}(\cdot , \psi))\, d\psi \, .
\end{gather}
The integral in the above equation can be approximated with a Riemann sum. Knowing the definition of GSWD, we define the objective function of GS-WGAN as follows:
\begin{gather}\label{equ: gswgan}
    \min_{\theta}  \max_{\phi} V_{GSW}(\pi_{\theta}, D_{\phi}) + \lambda R(D_{\phi}),\\
    \text{s.t.}\quad V_\mathrm{GSW}(\pi_{\theta}, D_{\phi}) = \int_{\psi \in \mathbb{S}^{d-1}} \mathbb{E}_{\mathbf{x}\sim p_\mathrm{r}}[\log D_{\phi}(\mathbf{x})] - \\
    \qquad \qquad \qquad \mathbb{E}_{\mathbf{y}\sim \pi_{\theta}}[\log D_{\phi}(\mathbf{y})] \, d\psi \, ,\nonumber
\end{gather}
where the parameters and notations are the same as defined in Eq.~\eqref{equ: wgan}. 

We note that $\mathbf{x}$ and $\mathbf{y}$ in Eq.~\eqref{equ: gswgan} are random variables in $\mathbb{R}^d$, which is not a reasonable assumption for graphs. To that end, we use an embedding function $g$ that maps each graph to a vector in $\mathbb{R}^d$. We use graph convolutional layers followed by fully connected layers to implement $g$. We deploy the same type of neural network architecture for $D_{\phi}$. We use $ \mathrm{R}_\mathrm{advers}=-V_\mathrm{GSW}(\pi_{\theta}, D_{\phi})$ as the adversarial reward used together with other rewards, and optimize the total rewards with a policy gradient method (Sec. 3.2.3).

\noindent\textbf{Network principle-based reward for drug combinations.}\quad Proteins or genes associated with a disease tend to form a localized neighborhood disease module rather than scattering randomly in the interactome \citep{cheng2019network}. A network-based score has been introduced \citep{menche2015uncovering}, to efficiently capture the network proximity of a drug ($X$) and disease ($Y$) based on the shortest-path length $d(x,y)$ between a drug target ($x$) and a disease protein ($y$):
\begin{equation}
    \begin{split}
        Z &= \frac{d(X,Y)- \bar{d}}{\sigma_d} \\
        d(X,Y) &= \frac{1}{||Y||} \sum_{y \in Y} \min_{x \in X} d(x,y),
    \end{split}
\end{equation}
where $d(\cdot,\cdot)$ is the shortest path distance; $\bar{d}$ and $\sigma_d$ are the mean and standard deviation of the reference distribution which is corresponding to 
the expected network topological distance  between two randomly selected groups of 
proteins matched to size and degree (connectivity) distribution as the original disease proteins and drug targets in the human interactome. Z-score being negative ($Z < 0$) implies network proximity of disease module and drug targets which is desirable. From the drug combination perspective, it has been shown that the complementary exposed drug-drug relationship has the least side drug side affect and the most drug combination efficacy \citep{cheng2019network}. Complementary exposed drug-drug ($X_1$ and $X_2$) relationship means that the drug targets ($x_1$) and drug targets ($x_2$) are not in the same neighborhood and has the least overlapping. Therefore, \citeauthor{cheng2019network} have proposed a network-separation score which is formulated as follow:
\begin{equation}
    s_{X_1,X_2} = d(X_1,X_2) - \frac{d(X_1,X_1) + d(X_2,X_2)}{2},
\end{equation}
where $d(X_1,X_2)$ is the mean shortest path distance between drugs $X_1$ and $X_2$; $d(X_1,X_1)$ and $d(X_2,X_2)$ are the mean shortest path distance within drug targets $X_1$ and $X_2$ respectively \citep{cheng2019network}. The separation score being positive ($s > 0$) implies to network are separated from each other which is desirable. \red{We} have extended and combined these scores for general drug combination therapy where we have a set of $k$ drugs $\{X_1,\cdots, X_k\}$ and disease $Y$:
\begin{equation}
    \mathrm{R}_\mathrm{network} = \lambda_1 \sum_{i=1}^{k} \sum_{j > i} s(X_i,X_j) - \lambda_2 \sum_{i=1}^{k} Z(X_i,Y)
\end{equation}

However, the exact online calculation of the reward $\mathrm{R}_\mathrm{network}$ is infeasible while training across all the diseases and the whole human interactome with more than 13K nodes and 352K edges. Therefore, we have developed a relaxed version of the reward which is feasible for online calculation and correlates with the actual reward. Specifically, we consider the normalized exclusive or (XOR) of intersections of disease modules with drug targets: 
\begin{equation}
    \hat{\mathrm{R}}_\mathrm{network} = \frac{Y \cap (X_1 \oplus \cdots \oplus X_k)}{|Y|} = \frac{(X_1 \cap Y) \oplus \cdots \oplus (X_k \cap Y)}{|Y|}.
\end{equation}
The relaxed network principle-based reward is penalizing a drug combination if the overlap between drug targets in the disease module is high, therefore it will prevent the adverse drug-drug interactions. \red{We scaled the network score by a constant (equals 10) such that the score would be in the same range as Pen-logP and can use the same weight in the total reward as Pen-logP did in \citep{you2018graph}.}

For a generated compound, we predict its protein targets by DeepAffinity \citep{karimi2019deepaffinity}, judging by whether the predicted IC$_{50}$ is below 1$\mu$M.   

\subsubsection{Policy Network}
Having explained the graph generation environment (various rewards), we outline the architecture of our proposed policy network. Our method takes the intermediate graph set $\mathcal{G}_t$ and the collection of scaffold subgraphs $C$ as inputs, and outputs the action $a_t$, which predicts a new link for each of the graphs in $\mathcal{G}_t$~\citep{you2018graph}.

Since the input to our policy network is a set of \red{$K$ compounds or} graphs $\{G_t^{(k)} \cup C\}_{k=1}^K$, we first deploy some layers of graph neural network to process each of the graphs. More specifically,
\begin{gather}
    X^{(k)} = \textsc{GNN}^{(k)} (G_t^{(k)} \cup C),\,\,\,\,\,\text{for}\,\,\, k=1,\dots,K,
\end{gather}
where $\textsc{GNN}^{(k)}$ is a multilayer graph neural network. The link prediction based action at iteration $t$ is a concatenation of four components for each of the $K$ graphs: selection of two nodes, prediction of edge type, and prediction of termination. Each component is sampled according to a predicted distribution \citep{you2018graph}. Details are included in the Supplemental Sec. 1.2.   
We note that the first node is always chosen from $\mathcal{G}_t$ while the \red{next} node is chosen from $\{G_t^{(k)} \cup C\}_{k=1}^K$. \red{We also note that}  infeasible actions (i.e. actions that do not pass valency check) proposed by the policy network are rejected and the state remains unchanged. We adopt Proximal Policy Optimization (PPO) \citep{schulman2017proximal}, one of the state-of-the-art policy gradient methods, to train the model.

\section{Results}

To assess the performance of our proposed model, we have designed a series of experiments. In section 4.1, we first compare HVGAE to state-of-art graph embedding methods in disease-disease network representation learning and further include several variants of HVGAE for ablation studies.  We then assess the performance of the proposed reinforcement learning method in two aspects. In a landscape assessment in Section 4.2, we examine designed pairwise compound-combinations for 299 diseases in quantitative scores of following a network-based principle \citep{cheng2019network}.  In Section 4.3, we focus on \red{four case studies involving multiple diseases of various systems-pharmacology strategies.  Our method is capable of generating  higher-order combinations of $K$ drugs.  As FDA-approved drug combinations are often pairs, here} we design \red{compound pairs} from the scaffolds of  FDA-approved drug \red{pairs}.  We further delve into designed compound \red{pairs} to \red{understand the benefit of following network principles in lowering toxicity from drug-drug interactions.  We also do so to} understand their systems pharmacology strategies in comparison to the FDA-approved drug  combinations.

\subsection{HVGAE representation compares favorably to baselines}

\subsubsection{Experiment setup }

To assess the performance of our proposed embedding method HVGAE, we compare its performance in (disease-disease) network reconstruction with Node2Vec \citep{grover2016node2vec}, DeepWalk \citep{perozzi2014deepwalk}, and VGAE \citep{kipf2016variational}, \red{as well as some variants of our own model for ablation study}. 
Node2Vec and DeepWalk are random walk based models that do not capture node attributes, hence we only used the disease-disease graph structure. For VGAE, we used identity matrix as node attributes as suggested by the authors. 

For our HVGAE described in Sec.~\ref{sec: hvgae}, we also considered two variants for ablation study: HVGAE-disjoint does not jointly embed gene-gene and disease-disease networks and does not use attentional pooling for disease embedding; whereas HVGAE-noAtt just does not use attentional pooling. Specifically, in HVGAE-disjoint, we, first, learned an embedding for gene-gene network, then used the sum of mean of the node representations of genes affected by a disease as its node attributes. In HVGAE-noAtt, we jointly learned the representations while using sum of mean of the node representations of genes as node attributes for disease-disease network. 

In node2vec and DeepWalk, the walk length was set to 80, the number of walks starting at each node was set to 10, and the nodes were embedded to a 16-dimensional space. The window size was 10 for node2vec while it is set to 10 in DeepWalk. All models were trained using Adam optimizer. In VGAE, a 32-dimensional graph convolutional (GC) layer followed by two 16-dimensional layers was used for mean and variance inference. The learning rate was set to 0.01. 

For HVGAE and its variants (for ablation study), we embed gene networks in 32 dimensional space using a single GC layer with 32 filters for each of the 5 types of input followed by a 64-dimensional GC layer and two 32-dimensional GC layer to infer mean and variance of the representation. We used a single 32-diensional fully connected (FC) layer for attention layer. For disease-disease network embedding, we deployed a single 32-dimensional GC layer followed by two 16-dimensional layer for mean and variance inference resulting in 16-dimensional embedding for disease-disease network. 
Learning rates were set to 0.001. The models were trained for 1,000 epochs choosing the best representation based on their the reconstruction performance at each epoch.

\subsubsection{Numerical analysis \red{and ablation study for network embedding}}
Table \ref{tab:auc_wna} summarizes the reconstruction performance of the aforementioned methods. \red{Compared to all baselines, our HVGAE showed the best performance in all metrics considered.}  Node2Vec and DeepWalk showed the worst performance as they only use the graph structure. The performance of VGAE was very close to DeepWalk. This is due to the fact that no attributes have been provided to VGAE despite having the capability of capturing attributes.

\begin{table}[!htb]
  \caption{Graph reconstruction performances  (unit: \%) in the disease-disease network  using our proposed HVGAE and baselines. F-1 scores are based on 50\% threshold.}
  \label{tab:results}
  \centering
  \resizebox{0.8\columnwidth}{!}{
  \begin{tabular}{@{}l | c c c c}
    \toprule
     {\textbf{Method}} & AUC-ROC & AP & F1-Macro & F1-Micro\\
    \midrule
    \midrule
    Node2Vec & 79.01 & 72.82 & 35.73 & 51.10\\
    DeepWalk & 79.32 & 73.77 & 40.28 & 53.30\\
    VGAE & 88.12 & 85.71 & 60.19 & 64.98\\
    \hline
    \textbf{HVGAE-disjoint} & 91.45 & 90.72 & 73.45 & 74.77\\
    \textbf{HVGAE-noAtt} & 92.83 & 92.34 & 73.81 & 75.14 \\
    \textbf{HVGAE} & $\mathbf{96.11}$ & $\mathbf{95.89}$ & $\mathbf{79.77}$ & $\mathbf{80.45}$\\
    \bottomrule
  \end{tabular}
  }
  \label{tab:auc_wna}\vspace{-0.2in}
\end{table}

\red{Compared to VGAE, HVGAE-disjoint without joint embedding or attentional pooling still saw better performance, which suggests that the attributes generated by the gene-gene network contains meaningful features about the disease-disease network. The slight performance gain from HVGAE-disjoint to HVGAE-noAtt shows that joint learning of both networks hierarchically helps to render more informative features for the disease-disease network. Finally, HVGAE had another performance boost compared to HVGAE-noAtt and outperformed all competing methods, which shows the benefit of attentional pooling. Specifically, the attention layer of HVGAE allows the model to produce features that are specifically informative for the disease-disease network representation learning.}

\subsection{Our model generates drug combinations following network principles across diseases}

\subsubsection{Experiment setup}
We have trained the proposed reinforcement model in 3 stages using different rewards, disease sets, and action spaces to increasingly focus on a target disease while exploiting all diseases who\red{se} representations already jointly embed gene-gene, disease-disease, and gene-disease networks. In the first stage, we train the model to only generate drug-like small-molecules which follow the chemistry valency reward, lipophilicity reward (logP where P is the octanol-water partition coefficient)~\citep{you2018graph}, and our novel adversarial reward for individual compounds. In this study, we trained the model for 3 days (4,800 iterations) to learn to follow the valency conditions and promote high logP for generated compounds. 

In the second stage, we start from the trained model at the end of the first stage (``warm-start'' or ``pre-training''). And we continue to train the model to generate good drug combinations across all diseases.  \red{We do so by} adding the network principle-based reward for compound combinations and sequentially generating drug combinations for each disease one by one. Then, we calculate the network-based score for the generated drug combinations at the last epoch across disease ontologies and compare them with the FDA-approved melanoma drug combinations' network-based score.  In this study,  we trained the model for 1,500 iterations to generate drug combinations across all 299 diseases.  In each iteration, we generated 8 drug combinations for a given disease. We adopted PPO \citep{schulman2017proximal} with a learning rate of 0.001 to train the proposed RL for both stages.

The last stage is disease-specific and will be detailed in Sec.~\ref{sec: cases}.

\subsubsection{Numerical analysis}
Across disease ontologies we quantify the performance of the proposed RL \red{(stage 2 model first)} using quantitative scores of compound-combinations following a network-based principle \citep{cheng2019network}. We consider the generated combinations in the last epoch (the last 299 iterations) and calculate the network score $\hat{\mathrm{R}}_\mathrm{network}$ based on disease ontologies. We asses our model based on two versions of disease classification, original disease ontology and its extension, explained in Sec.~\ref{sec:diseaseClass}. Table \ref{tab:nt_sc} summarizes the network-based scores for our model. Specifically, suppose that the set of targets for drug 1 and 2 are represented by $A$ and \red{$B$} whereas the disease module is the universal set $\Omega$, we report the portion exclusively covered by drug 1 ($\eta_{A-B}$), exclusively covered by drug 2 ($\eta_{B-A}$), overlapped by both  ($\eta_{A\cap B}$), and collectively by both ($\eta_{A\cup B}$).  As a reference, we calculated the corresponding network scores for 3 FDA-approved drug combinations for melanoma.

Based on the results shown in Table~\ref{tab:nt_sc}, we note that across all disease classes, the designed compound combinations learned in an environment, where the network principle\citep{cheng2019network} was rewarded, did achieve the desired performances. Specifically, their overlaps in disease modules were low as $\eta_{A\cap B}$ fractions are around 0.1; whereas their joint coverage in disease modules was high as $\eta_{A\cup B}$ fractions were in the range of 0.4--0.5 for all diseases.  

\begin{table}[!htb]
    \centering
    \caption{Network-based score for the generated drug combinations based on disease ontology classifications.}
    \resizebox{0.99\columnwidth}{!}{
   \begin{tabular}{@{}l | r r r r | r r r r}
    \hline
    \toprule
      & \multicolumn{4}{c|}{\textbf{Disease Ontology}}     & \multicolumn{4}{c}{\textbf{Disease Ontology extended}} \\
      & $\eta_{A-B}$  & $\eta_{B-A}$ & $\eta_{A\cap B}$ & $\eta_{A\cup B}$ & $\eta_{A-B}$  & $\eta_{B-A}$ & $\eta_{A\cap B}$ & $\eta_{A\cup B}$ \\
     \midrule
         infectious disease & 0.25& 0.10& 0.06& 0.41&0.20 & 0.07&0.05 &0.33 \\
         \hline
         disease of anatomical entity & 0.27& 0.12& 0.10& 0.49&0.26 &0.11 &0.09 &0.48 \\
         \hline
         disease of cellular proliferation & 0.25& 0.09& 0.07& 0.42&0.25 &0.10 &0.08 &0.44 \\
         \hline
         disease of mental health &0.22 &0.11 &0.10 &0.43&0.22 &0.11 &0.10 &0.43 \\
         \hline
         disease of metabolism & 0.22& 0.13& 0.10&0.46 &0.23 &0.14 &0.11 &0.48 \\
         \hline
         genetic disease &0.23 &0.15 &0.11 &0.4&0.23 &0.15 &0.11 &0.49 \\
         \hline
          	syndrome &0.22&0.11 &0.11 &0.44& 0.22&0.11 &0.11 &0.44 \\
         \hline
    \end{tabular}
    }
    \label{tab:nt_sc}
\end{table}

Compared to a few FDA-approved drugs for melanoma in Table 3, we notice that the designed compound combinations had similar exclusive coverage ($\eta_{A-B}$ and $\eta_{B-A}$) as the drug combinations.  However, the overlapping and overall coverage ($\eta_{A\cap B}$ and $\eta_{A\cup B}$) were both much higher in FDA-approved drug combinations than the designed.   Improvements could be made by training the RL agent longer, as these scores had  already been improving during the limited training process under computational restrictions.  More improvement can be made by adjusting the network-based reward as well. 

\begin{table}[!htb]
    \centering
    \caption{Network-based scores for  FDA-approved melanoma drug-combinations.}
    \resizebox{0.7\columnwidth}{!}{
   \begin{tabular}{@{}l | r r r r}
    \hline
      & $\eta_{A-B}$  & $\eta_{B-A}$ & $\eta_{A\cap B}$ & $\eta_{A\cup B}$  \\
     \midrule
         \hline
         Dabrafenib + Trametinib & 0.05&0.21 &0.55 &0.81 \\
         \hline
         Encorafenib + Binimetinib & 0.21& 0.05&0.53 &0.86 \\
         \hline
         Vemurafenib + Cobimetinib &0.05 & 0.27& 0.36& 0.68 \\
         \hline
    \end{tabular}
    }
    \label{tab:nt_sc_real}
\end{table}

\subsection{Case studies \red{for specific diseases}}\label{sec: cases}

\subsubsection{Experiment Setup}\label{sec:case-diseases}

In the third and last stage of RL model training, we start from the stage 2 model and generate drug combinations for a fixed target disease and can choose scaffold libraries specific to the disease.  \red{In parallel, we trained the model for 500 iterations (roughly 1 day) to generate 4,000 drug combinations specifically for each of 4 diseases featuring various  drug-combination strategies: melanoma, lung cancer, ovarian cancer, and breast cancer. In all cases, we started with the Murcko scaffolds of specific FDA-approved drug combinations to be detailed next.} 

\paragraph{Melanoma: Different targets in the same pathway.} Resistance to BRAF kinase inhibitors is associated with reactivation of the mitogen-activated protein kinase (MAPK) pathway. There is, thus, a phase 1 and 2 trial of combined treatment with Dabrafenib, a selective BRAF inhibitor, and Trametinib, a selective MAPK kinase (MEK) inhibitor.  As melanoma is not one of the 299 diseases, we chose broader neoplasm as an alternative.  To compensate the loss of focus on target disease, we design compound pairs from Murcko scaffolds of Dabrafenib + Trametinib.

\red{
\paragraph{Lung and ovarian cancers: Targeting parallel pathways.} MAPK and PI3K signaling pathways are parallels important for treating many cancers including lung and ovarian cancers~\citep{day2016approaches,bedard2015phase}. Clinical data suggest that dual blockade of these parallel pathways has  synergistic effects. Buparlisib (BKM120) and Trametinib (GSK1120212; Mekinist) are as a drug combination therapy are used for the purpose. Specifically, Buparlisib is a potent and highly specific PI3K inhibitor, whereas Trametinib is a highly selective, allosteric inhibitor of MEK1/MEK2 activation and kinase activity~\citep{bedard2015phase}.  
}

\red{
\paragraph{Breast cancer: Reverse resistance.} Endocrine therapies, including  Fulvestrant, are the main treatment for hormone receptor–positive breast cancers (80\% of breast cancers)~\citep{turner2015palbociclib}. However, they could confer resistance to patients during or after the treatment. 
A phase 3 study is using Fulvestrant and Palbociclib as a combination therapy to reverse the resistance. Fulvestrant and Palbociclib are targeting different genes in different pathways.  Specifically, Fulvestrant targets estrogen receptor (ER) $\alpha$ in estrogen signaling pathway and Palbociclib targets cyclin-dependent kinases 4 and 6 (CDK4 and CDK6) in cell cycle pathway~\citep{turner2015palbociclib}. 
}

\subsubsection{\red{Baseline methods for drug pair combination}}
\red{Since our proposed method is the first to generate drug combinations for specific diseases, we consider the following baseline methods to compare with: 1) random selection of 1,000 pairs from 8,724 small-molecule drugs in Drugbank \citep{wishart2018drugbank}; 2) 628 FDA-approved drug combinations curated by \citep{cheng2019network} for hypertension and cancers (our case studies are on 4 types of cancers); 3) random selection of 1,000 pairs of FDA-approved drugs for the given disease, based on drug-disease dataset "SCMFDD-L" \citep{zhang2018predicting}.}

\subsubsection{\red{Designed pairs follow network principles and improve toxicity}}

\begin{figure*}[!htb]
    \centering
    \includegraphics[width=1.0\textwidth]{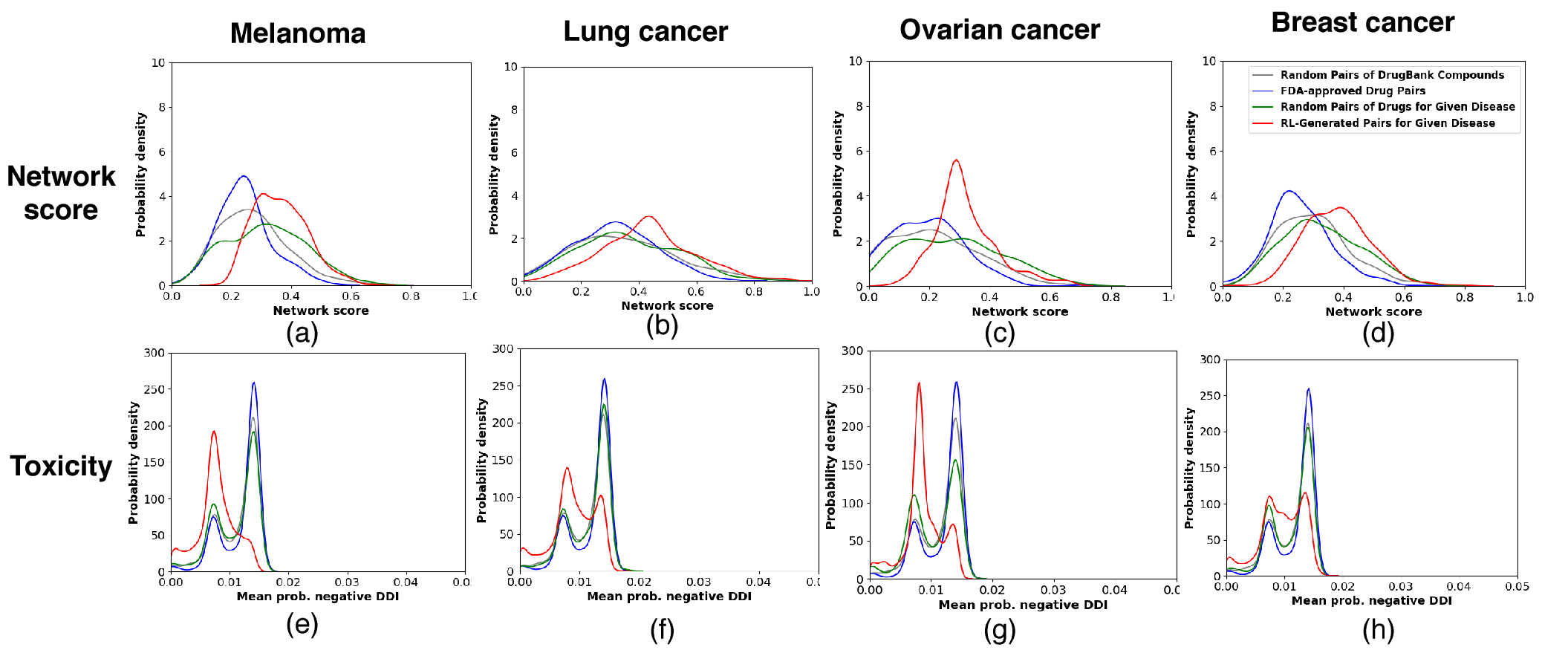}
    \caption{\red{Comparison of network score and toxicity of RL-generated pairs of compounds (our proposed method) with three baselines, i.e. random pairs of DrugBank compounds, FDA-approved drug pairs, and random pairs of FDA-approved drugs for four case-study diseases.}}
    \label{fig:result-net-tox}
\end{figure*}

\red{
We first compare the compound combinations designed by our model and those from the baselines using the network score that reflects the network-based principle.  Fig.~\ref{fig:result-net-tox}(a)--(d) shows that our designed combinations in all 4 cases, with higher network scores in distribution, respected the network principle more than the baselines (including the FDA-approved pairs not necessarily specific for the target disease). The observation is statistically significant with P-values ranging from 6E-74 to 7E-7 (one-sided Kolmogorov-Smirnov [KS] test; see more details in the Supplemental Tables S2 and S3).  Such a result is thanks to the network-principled reward we introduced.
}

\red{
We also examine whether drug combinations designed to follow the network principle could reduce toxicity from drug-drug interactions (DDIs).  DDIs are crucial when using drug combinations since they may trigger unexpected pharmacological effects, including adverse drug events (ADEs).  We used a deep-learning model DeepDDI \citep{ryu2018deep} with a mean accuracy of 92.4\% to predict for each combination the probabilities of 86 types of DDIs (we manually split them into 16 positive and 70 negatives; see details in the Supplemental Sec. 1.3).  To summarize over the DDIs, we considered both maximum and mean probabilities of positive or negative ones.  And we compared  those distributions between our designed pairs and baselines in each disease. }

\red{
Fig.~\ref{fig:result-net-tox}(e)--(h), using the mean probability among negative DDIs, shows that our compound pairs designed for all 4 diseases were predicted to have less chances of toxicity compared to the baselines.  One-sided KS tests attested to the statistical significance of the observation as P-values ranged between 2E-166 and 2E-53.  More analyses can be found in the Supplemental Sec. 3. 
}

\red{
Taken together, Fig.~\ref{fig:result-net-tox} suggested that following the network principle in designing drug combinations would help reduce toxicity due to DDIs.  
}

\subsubsection{\red{Designed pairs reproduce approved polyphamacology strategies}}

\red{
We next examine the DeepAffinity-predicted target genes of our designed pairs and compare them to the polyphamacology strategies outlined in Sec.~\ref{sec:case-diseases} for each disease. Since improved network scores  have been shown to correlate with lower toxicity, we used the scores to filter the 4,000 combinations designed for each disease. specifically, we retained combinations with network scores above 0.5 and $\eta_{A\cap B}$ below 0.1.  These designs are shared along with the codes.
}

\red{For melanoma, out of 69 combination designs retained, 26\% were predicted to jointly cover BRAF and MEK genes in a complementary way.} In other words, one molecule only target\red{s} BRAF and the other only target\red{s} MEK, according to our DeepAffinity\citep{karimi2019deepaffinity}-predicted IC$_{50}$, echoing the systems pharmacology strategy of the drug combination of Dabrafenib and Trametinib.  There were also other designs which demand further examination and potentially contain novel strategies. \red{All retained designs were predicted to target the MAPK pathway to which BRAF and MEK belong.}

\red{
For lung and ovarian cancers, the same filtering criteria retained 204 (896) compound combinations designed for lung (ovarian) cancer. As disease modules can be limited, MEK1/2 does not exist in the used modules for lung (ovarian cancer) and a gene-level analysis cannot be performed as the melanoma case.  Instead, we performed the pathway-level analysis and found that 50.9\% (45.2\%) of  combination designs for ovarian (lung) cancer were predicted to jointly and complementarily cover the MAPK and PI3K signaling pathways, which echoes the combination of Buparlisib and Trametinib.  Moreover, 99.5\% (100\%) of these retained designs were predicted to jointly target both pathways for  ovarian (lung) cancer. 
}

\red{
For breast cancer, 77 designed compound-combinations passed the filters. As CDK4/6 does not belong to the breast-cancer module due to the limitation of disease modules used, we again only performed a pathway-level analysis.  9\% of the combinations were predicted to jointly and complementarily cover ER-signaling and cell-cycle pathways as Fulvestrant and Palbociclib do.  Also, 74\% of the retained combinations jointly cover these pathways. These two portions suggest that many designed combinations were predicted to simultaneously target both pathways (with possible overlapping genes).  If we consider PI3K signaling  rather than cell cycle pathway for CDK4/6, 15.5\% of retained drug combinations were predicted to jointly and complementarily cover estrogen and PI3K signaling pathways and all of them did jointly.
}

\subsubsection{\red{Ablation study for RL-based drug-combination generation}}

\begin{figure}[!htb]
    \centering
    \includegraphics[width=0.4\textwidth]{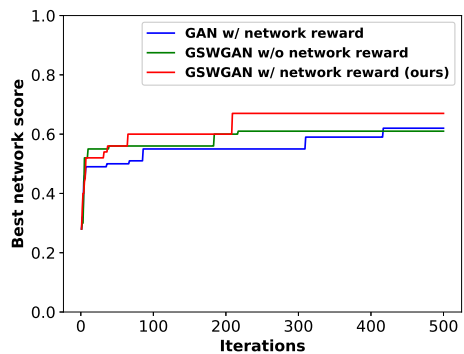}
    \caption{\red{Ablation study for RL: Best network scores achieved by three variants of the proposed method over training iterations.}}
    \label{fig:rl_ablation}
\end{figure}

\red{Besides HVGAE for network and disease embedding, two of our novel  contributions in RL-based drug set generations were network-principled reward  and adversarial reward through GS-WGAN. To assess the effects of these  contributions to our model, we performed ablation study for stage 3 using the case of melanoma. We ablated the originally proposed model in two ways:  removing the network-principled reward or replacing the GS-WGAN adversarial reward with the previously-used GAN reward based on Jenson-Shannon (JS) divergence. Results in Fig.~\ref{fig:rl_ablation} suggested that both rewards 
led to faster initial growth and higher saturation values in network-based scores.}

\section{Conclusion}
In response to the need of accelerated and principled drug-combination design, we have recast the problem as graph set generation in a chemically and net-biologically valid environment and developed the first deep generative model with novel adversarial award and drug-combination award in reinforcement learning for the purpose.  We have also designed hierarchical variation graph auto-encoders (HGVAE) to jointly embed domain knowledge such as gene-gene, disease-disease, gene-disease networks and learn disease representations to be conditioned on in the generative model for disease-specific drug combination.  Our results indicate that HGVAE learns integrative gene and disease representations that are much more generalizable and informative than state-of-the-art graph unsupervised-learning methods.  The results also indicate that the reinforcement learning model learns to generate drug combinations following a network-based principle thanks to our adversarial and drug-combination rewards.  Case studies involving four diseases \red{indicate that drug combinations designed to follow network principles tend to have low toxicity from drug-drug interactions. These designs also encode} systems pharmacology strategies echoing FDA-approved drug \red{combinations} as well as \red{other potentially promising strategies}.  As the first generative model for disease-specific drug combination design, our study allows for assessing and following network-based mechanistic hypotheses in efficiently searching the chemical combinatorial space and \red{effectively} designing drug combinations.  

\section*{Acknowledgements}
Part of the computing time is provided by the Texas A\&M High Performance Research Computing.   

\section*{Funding}
This project is in part supported by the National Institute of General Medical Sciences of the National Institutes of Health  (R35GM124952 to YS).

\bibliographystyle{natbib}

\bibliography{document}

\end{document}